\documentclass[10pt,leqno,twoside]{article}

\usepackage{annal-latex}
\usepackage[table]{xcolor}
\usepackage{graphicx}

\begin{document}
\setcounter{page}{1}
\newcommand\balline{\small Arthur-Jozsef Molnar}
\newcommand\jobbline{\small A Heuristic Process for Widget Matching}

\vspace{-4cm} \fofej{--}{--}{nn}{nnn}

\vspace{.4cm}

\title{A Heuristic Process for GUI Widget Matching Across Application Versions}

\author{{\bf Arthur-Jozsef Molnar} (Cluj-Napoca, Romania)\\[1ex]}

\mathclass{68N30,68N99}

\compclass{D2.5}

\projsupport{The author was supported by programs co-financed by The Sectoral Operational Programme Human Resources Development, Contract POS DRU 6/1.5/S/3 - ``Doctoral studies: through science towards society''}

\abstract {This paper introduces an automated heuristic process able to achieve high accuracy when matching graphical user interface widgets across multiple versions of a target application. The proposed implementation is flexible as it allows full customization of the process and easy integration with existing tools for long term graphical user interface test case maintenance, software visualization and analysis.}

\section{Introduction}
A large portion of modern software employs graphical user interfaces (GUIs) for interacting with users. A paradigm that has proven highly successful, GUIs are widely used in applications targeted for many types of devices such as notebooks, tablets, smartphones and the like. Technically, GUI driven software is a class of event-driven system where the GUI creates and fires events handled by the program underneath; it is this sequence of events that guides the application's execution.

Many modern applications expose their features through complex GUIs using multiple windows and hundreds of widgets to improve application presentability and ease of use. In many cases, GUI code can comprise more than half of the application code \cite{20}; being interacted with directly by end users, its correct functioning is considered crucial \cite{117}. 

During program execution the user interface is the only \emph{visible} component to users. However, many times its building blocks are found scattered across the application: windows and widgets are built using several approaches (e.g: writing source code, using design tools or declarative HTML or XML based technologies) while additional libraries can provide specialized components or display schemes\footnote{e.g: Java look\&feel}. These peculiarities demand new approaches in the design, visualization and testing of GUI driven software. 

This paper aims to address some of the existing challenges using an automated heuristic process able to match GUI widgets across application versions. This allows rapid creation of GUI test cases automatically reusable for new application versions. In addition, user interfaces can be compared between versions to provide important information about GUI changes that can be used in software understanding, visualization and testing.

This paper is structured as follows: the following section presents required preliminaries while the third section details the heuristic process. The next sections describe the implemented algorithms and our extensive case study. The last section is reserved for conclusions and future work planned.

\section{Preliminaries}
The main goal of our process is to improve automated GUI testing by reusing existing test cases across many application versions. This present section is dedicated to presenting the state of the art in GUI testing focused on previous work sitting at the foundation of our heuristic process.

GUI testing tools can be generally divided into two categories: capture-replay and model-based. Capture-replay tools are used in two distinct phases: during the \emph{capture} phase the program records the user's interactions with the target GUI and creates a recorded test case. The second phase consists of \emph{replaying} the recorded test case. The efficiency of these tools is related to how they record GUI interactions (e.g: do they use screen coordinates or they identify widgets using reflection or other means) and their capability at evaluating the result of running a test case. Adamoli et al. showed that capture-replay can be successfully used for automated performance testing \cite{120,121}, or for testing complex applications when combined with model based techniques \cite{122}.

The most important limitation of capture-replay derives from the lack of a backing model, which makes identifying and actioning GUI elements error-prone. This translates to uncertainty in recording and analyzing test results. These issues pushed recent date efforts towards model-based implementations. The existence of a formal model allows automatically building test suites, helps with evaluating test results and eases the implementation of visualization tools. 

Our research is based on the theoretical foundation provided in Memon's Phd. thesis \cite{20}, which defines the class of targeted GUIs as \emph{a hierarchical, graphical front-end to a software system that accepts as input user-generated and system-generated events, from a fixed set of events and produces deterministic graphical output. A GUI contains graphical objects; each object has a fixed set of properties. At any time during the execution of the GUI, these properties have discrete values, the set of which constitutes the state of the GUI.\cite{20}}. As a natural consequence, a GUI's state is represented using a set of objects and their associated properties. Our implementation represents the GUI as a tree of maps: each GUI object (or widget) is represented as a Map in which keys represent relevant properties and associated values provide the data (e.g: \emph{Class}="JButton", \emph{Text}="Ok", \emph{Background}="LightGray"). The GUI hierarchy is closely reflected in our implementation: the root node represents the GUI itself, its children represent application windows while widgets contained in windows start at the third level of the tree. Of course, not all widgets provide values for all properties. For example, only windows will have a \emph{Title} property, while widgets such as buttons and menu items will have \emph{Accelerator}, \emph{Text} or \emph{Icon} properties.

Many of the ideas detailed in \cite{20} were implemented in the GUITAR testing framework \cite{5,96}. A mature model-based implementation, GUITAR provides support for most of the major steps in testing: obtaining the GUI model, building and running test cases. Its maturity and consistency make it state of the art in GUI application testing. GUITAR consists of four applications:
\begin{itemize}
\item\emph{GUIRipper} automatically records the target application's GUI model. It employs reflection and automated interaction to record all accessible application windows, widgets and their properties, saving the resulting model in XML format \cite{95}. The current version supports Java, Windows and several mobile target platforms (Android and iOS).
\item\emph{GUI2EFG} computes the valid event sequences for a model obtained using GUIRipper. This is particularly important when building test cases, as not every widget is actionable at all times.
\item\emph{TestCaseGenerator} is used for building GUI test cases using the event-flow-graph built by \emph{GUI2EFG} and an existing GUI model. This tool supports plugins to allow for the implementation of various testing strategies.
\item\emph{TestCaseReplayer} is used to run the generated test cases and record their outcome. The tool saves the target application GUI state after each step, allowing detailed analyses to be performed after test suite execution.
\end{itemize}
However, GUITAR suffers from some of the problems outlined in the first section: changing an application's GUI renders many test cases useless due to changes in the model. This makes regression testing unfeasible as time invested for building compelling test suites cannot be redeemed by reusing them over many application versions.

These challenges have not gone unnoticed and several approaches have been proposed. An automated mechanism for repairing GUI test cases is presented in \cite{97}. It works by examining test cases that become unexecutable and repairing them by inserting or removing events. In \cite{98}, Huang et al. detail an approach for increasing test suite coverage by replacing unfeasible test cases with similar ones obtained using a genetic algorithm; they evaluate their implementation on a set of synthetic programs. These approaches prove that GUI test cases can be successfully repaired to run on newer versions of the target software. However they shift focus from exactly replaying older test cases which precludes their use for regression testing.

Preliminary work on GUI widget matching was presented by McMaster and Memon \cite{113}. Given two GUIs \emph{G} and \emph{G`}, they define the problem of matching equivalent widgets:

\emph{"For each actionable GUI element $e_i \in G$, find a corresponding element $e_j \in G'$ whose actions implement the same functionality." \cite{113}}
More formally, each widget from \emph{G} and \emph{G'} must be assigned to one of the following data structures \cite{113}:
\begin{itemize}
\item\emph{Deleted} - Contains elements only found in the \emph{older} GUI.
\item\emph{Created} - Contains elements only found in the \emph{newer} GUI.
\item\emph{Maintained} - This is a set of GUI element mappings of the type ($e_i \mapsto e_j$) that contains pairs of equivalent elements, with $e_i\in \emph{G}$ and $e_j\in \emph{G'}$
\end{itemize}
The efficiency of their proposed heuristic set is then examined using a typical \emph{Find/Replace} window.

\section{The Process}
This section details our implemented process based on the preliminaries above. The implementation comes in the form of a Java library designed for easy integration with other tools. To run the heuristic process the \emph{execute} method of the \emph{DefaultHeuristicService} class must be called using the following parameters:
\begin{itemize}
\item\emph{The GUIs}. The first two parameters encode the GUI models to be matched. To decouple our implementation from existing tools we implemented our own model, based on the one used by the GUITAR framework. A model transformer is available to enable transforming GUI models from GUIRipper format to our implementation; to use other GUI models, a new transformer implementation must be provided. 
\item\emph{A configuration file.} Using an XML configuration file allows flexibility in implementing custom heuristics and controlling the execution strategy.
\end{itemize}
The matching process works in three phases:
\begin{enumerate}
\item\emph{Window Matching.} The first phase of the process is matching the application windows. This is the only point where windows are matched, therefore the accuracy of the entire process is sensitive to errors in this step.
\item\emph{Widget Matching.} All GUI matches are performed during this phase\footnote{except windows as they are already matched}. This step is performed by running the widget matching heuristics over the window pairs matched in the previous step until no new matches are found.
\item\emph{Finalization.} This is the last step of the process. Widgets that remained unmatched are now classified as \emph{Deleted} or \emph{Created} based on the GUI version they belong to. This is a clean-up step that categorizes all previously unmatched elements.
\end{enumerate}
It is important to note that our process is able to match widgets only when they belong to matched application windows. This emphasizes the importance of correctly identifying matching windows and introduces limitations in identifying widgets moved across windows.

The matching process is customizable by providing a configuration file that provides the heuristic set to be used together with an execution strategy. The heuristics are provided as a prioritized list ordered by decreasing heuristic accuracy. The execution strategy is responsible with running the provided heuristics in a way that achieves maximum accuracy. The current implementation provides the following strategies:
\begin{itemize}
\item\emph{Cyclic Execution Strategy.} All heuristics are executed in order of priority and set to perform at most one match per execution. The process ends when running the full heuristic set reveals no new matches.
\item\emph{Priority Execution Strategy.} This strategy attempts to ensure that matches are always performed by the highest accuracy heuristic. To achieve this, the heuristics are executed in descending order of accuracy always restarting from highest accuracy whenever a match is found. This strategy has the effect that matches found by lower accuracy heuristics can be used by high accuracy implementations in finding new matching elements.
\end{itemize}
Our experiments showed priority execution strategy to be the most accurate. Custom execution strategies can be implemented by extending the \emph{AbstractHeuristicExecutionStrategy} class and providing a custom configuration file when running the process.
Existing heuristics are presented in the following section which also describes how new heuristics can be implemented and used in the process of matching both GUI windows and widgets.

\section{The Heuristics}
Our experimentation led to a number of heuristic implementations described in the following paragraphs, starting with the \emph{WindowMatchingHeuristic}, which is currently our only window matcher.

\subsection*{WindowMatchingHeuristic\footnote{All heuristics are implemented using Java classes of the same name}} This heuristic is used for matching application windows. Its accuracy is crucial as errors in window matching lead to none of the contained widgets being correctly paired. The implementation uses window titles to first match root windows and then to examine remaining window pairs. If both GUI versions have only one window they are matched regardless of title.

\subsection*{PropertyValuesHeuristic} This is an heuristic factory able to generate instances based on a provided configuration that defines the way property values must match according to provided criteria. The current implementation supports three criteria:
\begin{itemize}
\item \emph{Equality.} Property values must be both non-null and equal.
\item \emph{Similarity.} Property values must be similar when diffed \cite{114}. An integer parameter gives the maximum number of \emph{add} or \emph{delete} diff operations allowed for the values to be considered similar. When employing this criteria, the total number of \emph{add} and \emph{delete} operations represents the match score. Eligible widget pairs are then matched according to lowest match score. This approach allows for flexibility when property values change across GUI versions.
\item \emph{Nullity.} Both property values must be null.
\end{itemize}
Using these criteria a large number of heuristics can be generated. Note that the number or type of criteria is not limited. We can easily create a heuristic that searches for widgets with equal \emph{Class} property value, similar values for \emph{Text} and \emph{Icon} properties and null \emph{Accelerator} value.

\subsection*{PropertyValuesHierarchyHeuristic} This is a heuristic factory extending PropertyValuesHeuristic. The difference is that this implementation creates heuristics that search for widget matches only among children of already matched components. This adds an additional constraint based on evidence suggesting that many times matching GUI containers contain matching children. The separate implementation also allows optimizing the matching code to work faster for complex GUIs.

\subsection*{SingletonComponentHeuristic} This heuristic factory produces instances that match components based on the uniqueness of a provided property value. This is distinct from the implementations above as it enforces matched property values to be unique across the analyzed windows. For example, an instance using the \emph{Class} property would match a widget with \emph{Class} value of \emph{javax.swing.JButton} from the older window with another JButton instance only if they are both the only JButton's on their respective windows. Note that the implementation also counts already matched widgets when establishing uniqueness. This particular implementation was arrived at by analyzing the accuracy of the process for the jEdit \cite{104} text editor. Several of its versions implement custom components, such as \emph{HistoryTextField} which prove difficult to match correctly using only property-value based implementations. 

\subsection*{InverseHierarchyHeuristic} This heuristic is useful for matching components that were extensively modified across GUI versions. Given component $\emph{A} \in GUI_{older}$ and component $\emph{B} \in GUI_{newer}$, \emph{A} matches \emph{B} according to this implementation if:
\begin{itemize}
\item All children of \emph{A} have a matching component that is a child of \emph{B}.
\item \emph{A} has the same non-zero number of children as \emph{B}.
\end{itemize}
Our experiments showed this heuristic to be very accurate in correctly identifying menu items altered by changing their \emph{Text} and moving them within the menu system.

\subsection*{FinalHeuristic} The last phase of the matching process is performed by this heuristic. Its role is to assign all unmatched widgets in the older and newer versions to the \emph{Deleted} and \emph{Created} sets respectively. This final phase of the matching process cannot be customized and is implemented to ascertain that all GUI components are classified.

Our process is designed to be easy to set up, use and improve. In addition to the proposed heuristics, custom implementations can be provided by extending one of the following classes: \emph{AbstractGlobalHeuristic} is the base class used for window matchers while \emph{AbstractWindowHeuristic} is employed for matching widgets. Both classes have a \emph{run} method to receive required parameters and to  contain the matching logic. In addition, extending current implementations is possible in order to provide highly customized heuristics.

The heuristics hereby described are used in the case study discussed in the following section. This list however is neither extensive or comprehensive; we believe heuristics that provide higher accuracy exist and one of our future goals is targeted towards finding them. However, the following section shows that current implementations provide a solid base for deploying a highly accurate process that is generally applicable and reasonably efficient with regards to memory and processing requirements.

\section{Case Study}
We present a case study aiming to examine the accuracy of our proposed heuristic process applied to real-world GUI driven software. We examine the result of running the process and devise a number of metrics to answer the following research questions: (1) How can we measure the accuracy of the matching process? (2) What is the optimum heuristic process configuration? (3) How accurate is the process when applied to complex GUI software with the aim of enabling long term test-case maintenance and software visualization? (4) What type of matching errors can be expected and how can we limit their number?

To answer these questions we conducted a case study using several versions of popular open-source software available at SourceForge. We targeted applications with complex GUIs and long development history reflected in their source repositories.

\subsection*{Target applications} This section describes the two studied Java applications.
\subsubsection*{FreeMind} FreeMind is a widely used mind-mapping application\cite{103}, with over 14 million downloads and an 87\% user rating on the SourceForge website\footnote{as of 26.11.2011}. For our experiment we downloaded weekly versions from the FreeMind CVS and disregarded versions without source code changes. The final repository consists of 13 distinct versions spanning between November 2000 (version 0.2.0) and September 2007 (version 0.8.0b). According to hosting statistics the application was downloaded over 3.8 million times during that period. CVS timestamps and relevant information about the complexity of the studied versions are provided in Table \ref{tab:FreeMindVersionHistory}.

\begin{table}[t]\footnotesize
\centering
\begin{tabular}{ | c | c | c | c | c | c | }
 \hline
Version & CVS Timestamp & Classes & LOC & Widgets & Windows\\ \hline
0.1.0 & 01.11.2000 & 77 & 3597 & 101 & 1 \\ \hline
0.2.0 & 01.12.2000 & 90 & 4101 & 106 & 1 \\ \hline
0.2.0 & 01.01.2001 & 106 & 4453 & 132 & 1 \\ \hline
0.3.1 & 01.04.2001 & 117 & 6608 & 127 & 1 \\ \hline
0.3.1 & 01.05.2001 & 121 & 7255 & 134 & 1 \\ \hline
0.3.1 & 01.06.2001 & 126 & 7502 & 136 & 1 \\ \hline
0.3.1 & 01.07.2001 & 127 & 7698 & 137 & 1 \\ \hline
0.4.0 & 01.08.2001 & 127 & 7708 & 137 & 1 \\ \hline
0.6.7 & 01.12.2003 & 175 & 11981 & 244 & 1 \\ \hline
0.6.7 & 01.01.2004 & 180 & 12302 & 251 & 1 \\ \hline
0.6.7 & 01.02.2004 & 182 & 12619 & 251 & 1 \\ \hline
0.6.7 & 01.03.2004 & 182 & 12651 & 251 & 1 \\ \hline
0.8.0 & 01.09.2007 & 544 & 65616 & 280 & 1 \\ \hline
\end{tabular}
\caption{Versions of FreeMind used}
\label{tab:FreeMindVersionHistory}
\end{table}

The number of widgets increased almost three-fold from 101 to 280, making FreeMind's GUI a good candidate for evaluating the matching process. An important aspect concerning the studied versions was our inability to properly record the \emph{Options} window using \emph{GUIRipper}, which was therefore not taken into account in this study.

\subsubsection*{jEdit} jEdit is a popular pluginable text editor\cite{104} with a user rating of 77\% and over 6.6 million downloads to date. For our study we downloaded 17 released versions spanning January 2000 (version 2.3pre2) to May 2010 (version 4.3.2final). SourceForge statistics reveal over 5.7 million downloads during that period. Table \ref{tab:jEditVersionHistory} presents the versions in our repository together with relevant information for our process.

\begin{table}[t]\footnotesize
\centering
\begin{tabular}{ | c | c | c | c | c | c | }
 \hline
Version & CVS Timestamp & Classes & LOC & Widgets & Windows \\ \hline
2.3pre2 & 29.01.2000 & 332 & 23709 & 482 & 12 \\ \hline
2.3final & 11.03.2000 & 347 & 25260 & 533 & 14 \\ \hline
2.4final & 23.04.2000 & 357 & 25951 & 559 & 14 \\ \hline
2.5pre5 & 05.06.2000 & 416 & 30949 & 699 & 16 \\ \hline
2.5final & 08.07.2000 & 418 & 31085 & 701 & 16 \\ \hline
2.6pre7 & 23.09.2000 & 456 & 35020 & 591 & 12 \\ \hline
2.6final & 04.11.2000 & 458 & 35544 & 600 & 12 \\ \hline
3.0final & 25.12.2000 & 352 & 44712 & 584 & 13 \\ \hline
3.1pre1 & 10.02.2001 & 361 & 45958 & 590 & 13 \\ \hline
3.1pre3 & 11.03.2001 & 361 & 46165 & 596 & 13 \\ \hline
3.1final & 22.04.2001 & 373 & 47136 & 648 & 13 \\ \hline
3.2final & 29.08.2001 & 430 & 53735 & 666 & 12 \\ \hline
4.0final & 12.04.2002 & 504 & 61918 & 736 & 13 \\ \hline
4.2pre2 & 30.05.2003 & 612 & 72759 & 772 & 13 \\ \hline
4.2final & 01.12.2004 & 650 & 81755 & 860 & 14 \\ \hline
4.3.0final & 23.12.2009 & 872 & 106398 & 992 & 16 \\ \hline
4.3.2final & 10.05.2010 & 872 & 106510 & 992 & 16 \\ \hline
\end{tabular}
\caption{Versions of jEdit used}
\label{tab:jEditVersionHistory}
\end{table}

Across the studied versions, the number of code lines increases from 23 thousand to over 100 thousand which is coupled with an effective doubling in widget count. Also, the number of windows targeted by our process varies between 12 to 16 according to application version. The only caveat regards jEdit's \emph{Options} window which could not be ripped correctly and was therefore excluded from our study.

\subsection*{Heuristic metrics} This section describes our efforts in answering research question (1). To the best of our knowledge the hereby proposed process is a first of its kind and therefore requires new metrics to evaluate its accuracy. The first step was to create oracles to provide the correct match decisions for every version pair studied. This was achieved manually with the aid of a widget comparison module added to our jSET tool \cite{116}, available on our SVN repository \cite{105}. This is how the 28 required oracles were obtained\footnote{We have 30 application versions which lead to 29 pairs; however we use 2 different applications and so we have 28 version pairs}.

We started by defining a few measurements available using the oracle data itself:
\begin{itemize}
\item \emph{Correct Decision Count (CDC).} The number of correct decisions for a given version pair. We define a decision as the action taken by the process to classify a widget as \emph{Deleted}, \emph{Created}, or a widget pair as \emph{Maintained}. Therefore this represents the total number of elements in the three defined data structures. Note that this differs from the number of widgets because a matched pair in the \emph{Maintained} structure is counted as one decision. CDC represents the number of correct decisions that have to be taken for maximum accuracy.
\item \emph{Correct Match Count (CMC).} The number of elements in the \emph{Maintained} structure. This represents how many elements in the older GUI have an equivalent in the newer version.
\item \emph{Dissimilar Widget Count (DWC).} Represents the number of widgets changed in the studied version pair. This includes all widgets in the \emph{Deleted} and \emph{Created} structures together with the number of widgets in matched pairs where at least one property value that does not refer to widget size or location has changed. As an example, if a button's icon is changed across versions the widgets are considered dissimilar. However, if only the position and size of the button changes it is not categorized as such.
\end{itemize}
After running the heuristic process our automated evaluation algorithm analyzes the obtained results alongside available oracle data and computes values for the following metrics:
\begin{itemize}
\item \emph{Heuristic Correct Decision Count (HCDC).} The number of correct decisions taken by the process. Its value is a number between 0 (no correct decisions taken) and the CDC value available using oracle data.
\item \emph{Heuristic Correct Match Count (HCMC).} The number of correctly determined functionally equivalent GUI element pairs. This is the number of elements correctly assigned to the \emph{Maintained} structure computed by the heuristic. Its value is a number between 0 (no correct matches) and the CMC value available using oracle data.
\item \emph{Heuristic Correct Decision in Dissimilar Widgets Count (HCDDWC).} The number of correct decisions taken for dissimilar widgets. Its value is a number between 0 and the DWC available using oracle data.
\end{itemize}
To better understand the accuracy of a heuristic run the following measurements are calculated:
\begin{itemize} 
\item \emph{Heuristic Decision Rate (HDR).} The percentage of correct heuristic decisions, calculated as $\frac{HCDC}{CDC}$.\footnote{While these measurements take values between 0 and 1 we use percentages for convenience.}
\item \emph{Heuristic Match Rate (HMR).} The percentage of correct heuristic matches, calculated as $\frac{HCMC}{CMC}$.
\item \emph{Heuristic Dissimilar Widgets Decision Rate (HDWDR).} The percentage of correct decisions for dissimilar widgets, calculated as $\frac{HCDDWC}{DWC}$.
\end{itemize}
Implementing multiple metrics allows us to better study the accuracy of different heuristic sets and to better ascertain how well the process is suited to different goals. The overall accuracy of the process is best expressed using the \emph{Heuristic Decision Rate}, as it takes all GUI elements into account. A high \emph{Heuristic Match Rate} is important for enabling GUI test-case maintenance because it shows that most equivalent widget pairs were correctly identified; this leads to test suites that have a long life across multiple application versions. The long named \emph{Heuristic Dissimilar Widgets Decision Rate} can assess the accuracy of the heuristic set for changed elements, which is important when analyzing frequently changing GUIs. Due to the particular nature of the problem we decided these measurements to have greater value than classical false positive/negative analyses. However, new metrics are easy to define and implement as our tool source code is available for download \cite{105}.

\subsection*{Best heuristic set} We performed several experiments in order to answer research question (2). As the process is implemented as a prioritized list of heuristics we ordered our implementations in descending order of accuracy. This ensures that matches are always determined by the most accurate heuristic available. Using our data repository and oracle information we prioritized widget properties according to how well they indicate component equivalency. Our experimentation showed several property prioritizations yielded good results, with differences usually within 1 percentage point when aggregated over the entire data set. The prioritized property table used in our case study is shown in Table \ref{tab:Priorities}. A script was implemented to generate a comprehensive set of heuristics starting from the provided data.

\begin{table}[t]\footnotesize
\centering
\begin{tabular}{ | c | c | }
\hline
\rowcolor{lightgray} Property & Meaning \\ \hline
Hierarchical & Matching widgets must have matching ancestors \\ \hline
(diff) Icon & Widget icon names are diffed to detect changed icons \\ \hline
Icon & Widget icon name \\ \hline
(diff) Text & Widget text is diffed to detect changed text \\ \hline
Class & Widget class \\ \hline
Text & Widget text \\ \hline
Accelerator & Widget keyboard accelerator \\ \hline
Index & Widget index in parent container \\ \hline
Width Height & Widget size \\ \hline
X Y & Widget location in window \\ \hline
\end{tabular}
\caption{Widget properties in descending order of heuristic accuracy for widget matching}
\label{tab:Priorities}
\end{table}

Heuristic implementations are generated by running two loops over the property list. The outer loop controls the number of disregarded properties ($n_{dp}$) and takes values from 0 to the number of properties minus 2. By creating the strictest heuristics first we preserve the inherent structure of the priority list. The "minus 2" limit was imposed to prevent the creation of permissive heuristics that fail to provide accurate matches. The second loop controls which properties are disregarded. This loop traverses the table from bottom to top disregarding $n_{dp}$ properties. At each step a new heuristic is generated and added to the priority list. For example, if all values but \emph{Hierarchical}, \emph{Icon} and \emph{diff Text} are disregarded (when $n_{dp}=7$ for the table above) a \emph{PropertyValuesHierarchyHeuristic}\footnote{when the \emph{Hierarchical} property is not disregarded hierarchical heuristics are generated} that checks for \emph{Icon} equality and \emph{Text} similarity is generated. Note that although Table \ref{tab:Priorities} has 10 rows, we have distinct rows for \emph{Text} and \emph{Icon} criteria. When a heuristic is generated and provided with multiple criteria for a given property the strictest one is used\footnote{In our case property value equality}. Another aspect relates to property grouping. In Table \ref{tab:Priorities} properties relating to widget location and size are grouped so generated heuristics evaluate them together. This makes sense because while we refer to a widget's location or size as a single logical property, they are actually expressed as distinct properties in our model.

We experimented with property order to find generally applicable configurations able to provide high accuracy. Our most accurate configuration is the one in Table \ref{tab:Priorities}. However we found that certain scenarios were not properly addressed using this approach. One of them consists of custom implementations of GUI components used across several versions (e.g: jEdit's \emph{HistoryTextField}) that were consistently mismatched. Another problem was frequent changes to some containers, notably application menus. Both our target applications have menu structures that were significantly altered across the studied versions. We believe these issues to be of a general nature not limited to our studied applications. The solution to these problems lay in implementing two additional high accuracy heuristics: \emph{SingletonComponentHeuristic} and \emph{InverseHierarchyHeuristic}, which solved the majority of the issues.

\subsection*{Case study results} This section addresses research question (3) by presenting results obtained when applying our heuristic process to the 28 version pairs of FreeMind and jEdit. These experiments were run using our test harness capable of automatically running and evaluating heuristic configurations.

The results below were obtained using the heuristic set detailed in the previous section, filtered by removing low performing implementations and to which \emph{SingletonComponentHeuristic} and \emph{InverseHierarchyHeuristic} instances were added. We believe this set provides a good balance between accuracy, generality and speed of execution.

During our experimentation we observed several factors that skewed the obtained results. The first such factor regards composite widgets. These are UI components that consist of multiple widgets working together to enable complex behaviour. A prime example is the combo-box, which usually consists of a text field, a button and a scrollable list. The second factor is the presence of certain components used for delimitation purposes, such as menu separators, invisible panels and such.

As our process targets test-case maintenance and software visualization we decided to ignore delimitation widgets in our evaluation and to disregard children of composite widgets. While we believe these settings provide the most accurate representation of the matching process results, changing ignored widget types (including children of composite widgets) is easily accomplished using our process'  configuration files.

Table \ref{tab:AggregateResults} shows the aggregate results obtained by our best heuristic set over the 28 version pairs. Note that the presented metrics were aggregated over the entire data set.

\begin{table}[t]\footnotesize
\centering
\begin{tabular}{ | l | c | c | c | }
\hline
Measurement & FreeMind & jEdit & Total \\ \hline
Correct Decision Count & 1799 & 8976 & 10775 \\ \hline
Correct Match Count & 1524 & 7461 & 8985 \\ \hline
Dissimilar Widget Count & 797 & 4115 & 4912 \\ \hline
Heuristic Decision Count & 1787 & 8953 & 10740 \\ \hline
Heuristic Match Count & 1505 & 7321 & 8826 \\ \hline
Heuristic Correct Decision Count & 1743 & 8436 & 10179 \\ \hline
Heuristic Correct Match Count & 1502 & 7194 & 8696 \\ \hline
\rowcolor{lightgray} Heuristic Decision Rate & 96.89\% & 93.98\% & 94.47\% \\ \hline
\rowcolor{lightgray} Heuristic Match Rate & 98.56\% & 96.42\% & 96.78\% \\ \hline
\rowcolor{lightgray} Heuristic Dissimilar Widget Decision Rate & 89.46\% & 80.46\% & 81.92\% \\ \hline
\end{tabular}
\caption{Heuristic process results}
\label{tab:AggregateResults}
\end{table}

The first observation regards the sheer size of the data set. The 28 version pairs studied contain close to 9000 equivalent widget pairs, more than half of which are considered dissimilar. It is worth noting that in our interpretation resizing or moving widgets onscreen does not cause them to be considered dissimilar, even though this would break many existing implementations for test case replay.

The most important results are found in the highlighted rows. Considering the time span of the targeted versions (7 years for FreeMind and 10 for jEdit) we consider decision rates well over 90\% as very promising. We believe match rates over 95\% enable the implementation of long-lived GUI test cases adaptable across multiple application versions. Also, a high dissimilar widget decision rate shows that the process is able to match heavily modified widgets successfully, making it feasible for implementation in quickly evolving applications.

An interesting observation is that our process was consistently more accurate for the FreeMind application. This was expected due to its simpler user interface and limited use of custom components.

\begin{table}[t]\footnotesize
\centering
\begin{tabular}{ | c | c | c | c | }
\hline
FreeMind & Decision Rate & Match Rate & Dissimilar Decision Rate \\ \hline
100\%     & 8 & 9 & 8 \\ \hline
above 95\% & 10 & 11 & 9 \\ \hline
above 90\% & 11 & 11 & 9 \\ \hline
above 80\% & 12 & 12 & 11 \\ \hline
\end{tabular}
\caption{FreeMind Result consistency}
\label{tab:FreeMindConsistence}
\end{table}

While the process obtained good aggregate results, we were also interested in achieving good consistency across the studied versions. Table \ref{tab:FreeMindConsistence} details the consistency results obtained for the 12 FreeMind version pairs. We observe that 8 version pairs exhibit flawless heuristic performance, and match rates are over 80\% for all studied pairs. The only disheartening result occurs in one version pair that did not achieve 80\% decision rate when examining modified widgets. Manual examination of process log files revealed that only 4 dissimilar widgets were found, and for 2 of them wrong decisions were taken. Table \ref{tab:jEditConsistence} provides information about the consistency of the heuristic process for the 16 jEdit version pairs. Due to increased complexity and longer time span elapsed between the tested versions we manage to reach 95\% decision accuracy for only half the studied version pairs. By comparing results' consistency it becomes clear that our process is feasible for long-term test-case maintenance and comparative GUI visualization when using short iterations.

\begin{table}[t]\footnotesize
\centering
\begin{tabular}{ | c | c | c | c | }
\hline
jEdit & Decision Rate & Match Rate & Dissimilar Decision Rate \\ \hline
100\%      &  0 &  1 & 0 \\ \hline
above 95\% &  8 & 12 & 0 \\ \hline
above 90\% & 14 & 16 & 0 \\ \hline
above 80\% & 16 & 16 & 9 \\ \hline
\end{tabular}
\caption{jEdit Result consistency}
\label{tab:jEditConsistence}
\end{table}

Our website \cite{105} contains all the required tools for duplicating this case study. Target application source code, recorded GUIs and pre-computed oracle information are readily available for all studied versions on our SourceForge SVN repository. The implementation of the heuristic process described in the previous sections is hosted at the same address and available under a popular free software license.

\subsection*{Heuristic error analysis} An important aspect for improving process accuracy regards the thorough analysis of the heuristic errors. This section attempts to provide an answer to research question (4) by analyzing some common types of mismatches that occurred in our case study. 

\subsubsection*{Detecting changes to complex widgets} Most GUIs contain widgets displaying complex behaviour such as combo-boxes or tables. We observed that when such components were changed across application version, especially by both resizing and moving they were sometimes incorrectly identified. This was due to the limited number of properties associated with these components. As such, a future direction is implementing extensible custom matching routines to target complex widget types.

\subsubsection*{Changes beyond GUI level} 
\begin{figure}[htbp]
	\centering
	\includegraphics[width=\textwidth]{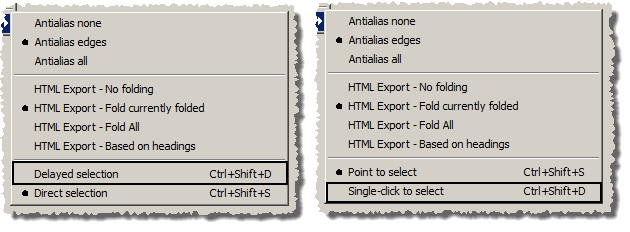}
	\caption{Highlighted items have the same accelerator but are not equivalent}
	\label{1}
\end{figure}

FreeMind's evolution between January and February 2004 brought a change in one of the menu items within the \emph{Edit} menu that could not be detected using GUI analysis. This was due to changes at both GUI and event handling level which required source-code analysis to correctly identify the matching widgets. Figure \ref{1} shows the menu items, with the heuristic decision highlighting the erroneously matched pair. Since our process does not leverage source code information it failed to take the correct decision.

\subsubsection*{Changes in widget actions} Some of the erroneous decisions taken when analyzing jEdit were due to changes in event handling that were only subtly reflected in the GUI. Such was the case of some menu items that maintained their place and accelerator in the GUI hierarchy with changes being evident by small modifications of the displayed text.

Our analysis enabled us to identify some areas for improvement. We believe some error types can be eliminated and future work will be focused on doing so. Special notice must be awarded however to changes that go beyond GUI level, as these can be detected only via source code analyzers such as the SPARK framework provided with Soot \cite{2}.

\subsection*{Threats to validity} While we tried to reduce threats to the validity of our case study results we identified some aspects that need further detailing. The first identified threat relates to generality and is based on the limited scope of our experimental targets. Our chosen target applications represent only a fraction of all possible implementations and as such might not be representative for all GUIs. We tried to mitigate this aspect by carefully choosing complex GUI driven applications that were employed in multiple previous studies \cite{99,118,119,120} and providing ample means of customizing the heuristic process. A second threat regards internal validity. Our process is fully automated and multiple software safeguards were implemented to ensure error-free operation. Also, matches obtained by the heuristic process were double checked with oracle data to eliminate possible errors in both. We believe our process is free from significant errors in this respect and as such to be suited for use across a wide range of applications.

\section{Limitations and Future Work}
Although our matching process was designed for maximum flexibility, we identified some aspects that might limit its use. Some of the aspects stem from the tools we based our research on, while others are good candidates for future efforts. The following list attempts an overview of these limitations:
\begin{itemize}
\item \emph{Dynamic user interfaces.} Some applications create and dispose of GUI elements dynamically; recording these would require using a specification language to describe the rules that govern GUI element creation and disposal, a task that brings added complexity to the process. Also, user interfaces that have timing issues (e.g: web interfaces) or that present a continuous stream of data (e.g: media players) cannot be completely captured by the GUIRipper tool \cite{20}.
\item \emph{Custom user components.} While GUIRipper can be considered a mature tool, it is not capable of fully recording every application's GUI. More so, some applications implement custom components or workarounds to platform-specific issues which preclude the recording of their properties. While our process was designed with this in mind, lack of data does lead to poor heuristic performance. For example, certain versions of jEdit implement custom menu items to provide key accelerators. These could not have been recorded and interpreted correctly without modifying the GUIRipper component.
\item \emph{Magnitude of changes.} We observed our process performed consistently better for the FreeMind application. The main reason was that we obtained the studied versions directly from source control, so GUI changes could not accumulate as they did in the case of jEdit. This leads us to believe that our process is best employed when target application versions are close to each other on the time scale.
\end{itemize}

We believe our process has extensive applications in GUI software testing and visualization, therefore our goal is to continually extend and improve its scope. A future direction consists of performing a more consistent case study by including several .NET and SWT based GUI applications. Such a study can provide answers relating to the generality of the heuristic process together with additional data about how the accuracy of the process changes when examining nightly or weekly application builds. In addition, this would allow better understanding of how accuracy is affected by custom widget implementations.

Another direction regards extending this research beyond the desktop paradigm by including web and mobile applications. Bryce el. al devised a common model for event driven software testing in \cite{1}. An elaborate study is required to learn whether the current process is feasible using a general model and to find suitable heuristic implementations that work for web and mobile applications.


\vspace{2cm}

\noindent\textbf{Arthur-Jozsef Molnar}\\
Department of Computer Science, Faculty of Mathematics and Computer Science, Babe\c{s}-Bolyai University\\
Cluj-Napoca\\
Romania\\
{\tt arthur@cs.ubbcluj.ro}\\

\end{document}